\documentclass[aps,prb,twocolumn]{revtex4}
\usepackage{epsfig}
\usepackage{color}
\usepackage{amsmath}
\usepackage{amssymb}
\usepackage{bbm}

\newcommand{\be}{\begin{equation}}
\newcommand{\ee}{\end{equation}}
\newcommand{\bea}{\begin{eqnarray}}
\newcommand{\eea}{\end{eqnarray}}

\begin{document}

\title{Impact of the electric field on superconductivity in Bose-Einstein Condensation regime.} 

\author{Naoum Karchev}

\affiliation{Department of Physics, University of Sofia, 1164 Sofia, Bulgaria}

\begin{abstract}

In the strong coupling Bose-Einstein Condensation (BEC) regime the superconductors have two characteristic temperatures: $T^*$- onset of fermion pairing and 
$T_{sc}$- onset of superconductivity, 
such that  $T_{sc}<T^*$. In the present article, we consider time dependent Ginzburg-Landau theory of superconductors with slab geometry and show that applied electric field, in the temperature interval $(T_{sc},T^*)$, Bose condenses the Cooper pairs thereby increasing 
the superconductor critical temperature $T^E_{sc}>T_{sc}$. Important consequence is the fact that arbitrary temperature within the interval ($T_{sc},T^*$) is a critical temperature of superconductor transition if an appropriate electric field is applied. 
This means that if we set the temperature of the system within the above mentioned interval and increase the applied electric field the system undergoes an electric field induced transition to superconductor.
 We also show the existence of critical value of the applied electric field at which $T^*=T^E_{sc}$. This means that although the system is in  BEC regime, 
 away from the BCS one, we can apply an electric field that moves the system to a state with $T^E_{sc}=T^*$, characteristic of BCS regime. 
The results indicate that applied electric field experiments are a suitable tool to identify the BEC regime of the superconductors. The experiment can determine 
$T^*$ as a temperature below which the  electric field Bose condenses the Cooper pairs, while above it the electrons screen the field and it cannot penetrate.

\end{abstract}

\maketitle

\begin{figure}[!ht]
	\epsfxsize=\linewidth
	\epsfbox{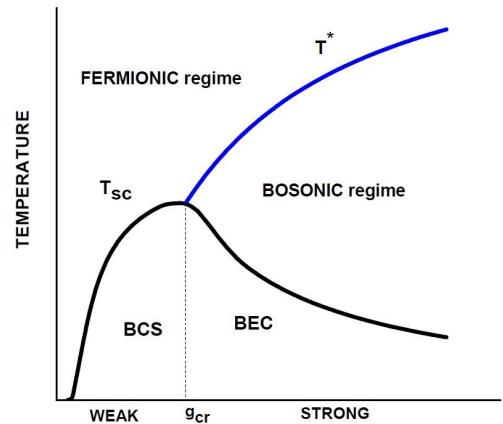} 
	\caption{\,\,  Schematic phase diagram of BCS-BEC crossover. The curve $T_{sc}$ is superconductor critical temperature as a function of attraction constant-$g$. The curve $T^*$ shows the critical temperature  of fermion pairing  as a function of $g$. $g_{cr}$ is the critical value of the BCS-BEC transition , although the crossover is characterized by interval rather then point. Weak coupling superconductivity is well explained by Bardeen, Cooper and Schrieffer (BCS) theory, while the strong coupling superconductivity rely on Bose Einstein Condensation (BEC) theory of fermion pairs formed  above $T_{sc}$ but below $T^*$ (BOSONIC regime). }\label{(fig1)E-BEC-Sc}
\end{figure}

The superconductivity is a consequence of attractive interaction between fermions. This causes  the pairing of fermions. The fluctuations of bosonic 
Cooper-pairs drive the system to Bose condense and to onset of superconductivity. The system has two characteristic temperatures: $T^*$-onset of fermion pairing and $T_{sc}$-onset of superconductivity. The temperature-attraction constant $g$ diagram is illustrated schematically in (Fig.\ref{(fig1)E-BEC-Sc}).

In the weak constant 
regime the electron pairs are large, overlapping and the superconductivity is well described by Bardeen, Cooper and Schrieffer (BCS) theory \cite{BCS57}.  
The critical temperature $T_{sc}$ is equal to $T^*$ ($T_{sc} = T^*$), and as a function of constant $g$ has a form $T_{sc}\propto e^{-const/g}$. 
In strong coupling regime the electron pairs are small, tightly bound, $T_{sc}<T^*$, which means that fermion pairs form before the onset of superconductivity. In temperature interval $(T_{sc},T^*)$ the Schafroth idea for superconductivity \cite{Schafroth55} is useful. He claims that the onset of superconductivity could be thought of 
as Bose condensation of ideal charged Bose gas, and the expression for the critical temperature follows that of an ideal Bose gas (BEC theory). There are theoretical arguments that system of Fermi pairs behaves as a Bose gas whose particles (bound states) form a condensation at low temperature \cite{Popov66}. 

The first experimental realization of BEC-BCS crossover was achieved in ultracold  atomic gases \cite{Regal03,Jochim03,Zwierlein03,Regal04}.
The BCS-BEC crossover has been extensively studied in condensed matter physics \cite{Eagles69,Laggett80,NS-R85,Ullah91,Chen98,Larkin02,Iddo02,Randeria89,Randeria90,Drechsler92,Melo93,Randeria97,Ioffe97,Benfatto01,Ricci10,Robin95,Levchenko11,Chen05,Uemura91}. 
Special efforts are made to establish the existence of Cooper pairs above critical temperature and in this way to verify the BCS-BEC crossover in the system. It was pointed out that the evolution of the system from weak to strong coupling superconductivity is smooth \cite{NS-R85}.

Of particular interest are 2D superconductors. The onset of superconductivity in 2D requires an existing of electron pairs \cite{Kosterlitz16}. 
In 2D superconductors BCS-BEC crossover is smooth \cite{Randeria89}, occurs at weak coupling and can be used weak coupling theory \cite{Chubukov19}. 
The likely existence of pre-formed pairs in two rather different materials, 
a high-temperature cuprate superconductor and strontium titanate, are highlighted recently \cite{Ivan20}. The experiments with underdoped copper oxide $Bi_2Sr_2CaCu_2O_8$ \cite{Yang08,Kondo11}
show that the pseudogap does reflect the formation of preformed pairs of electrons.
Crossover behavior from the BCS limit to the BEC limit realized by varying carrier density in a two-dimensional superconductor, electron-doped zirconium nitride chloride is reported in \cite{Nakagawa21}. The small size of cuprate pairs is an additional argument supporting BCS-BEC scenario \cite{Leggett06}.

Contrary to the above arguments there are claims \cite{Tallon99} suggesting that quantum critical behavior is presented so that $T^*$ actually vanishes under the superconducting dome. This is inconsistent with a BCS-BEC picture in which $T^*$ is necessarily larger than $T_{sc}$.

There are a number of signatures of BCS-BEC crossover \cite{Chen05}. Controversies in the interpretation of experimental results indicate 
that a new type of experiment is needed. We propose to study superconductors by applying an
electric field. In the present paper we show that applying electric field on superconductors in Bose-Einstein Condensation regime below $T^*$ the Cooper pairs Bose condense and the superconducting critical temperature $T_{sc}$ increases. This result can serve as proof of the existence of fermion pairing above the critical temperature, i.e for the existence of BEC regime. An arbitrary temperature within the interval ($T_{sc},T^*$) is a critical temperature of superconductor transition if an appropriate electric field is applied. This means that if we set the temperature of the system within the interval $(T_{sc},T^*)$ and increase the applied electric field the system undergoes an electric field induced transition to superconductor. 
 We prove that if the system is in  BEC regime, 
away from the BCS one, we can apply an electric field that moves the system to a state with $T^E_{sc}=T^*$, characteristic of BCS regime. 

To begin with we consider time dependent Ginzburg -Landau theory \cite{GL50,Gorkov68,Thompson70,Tinkham75}, written in terms of gauge four-vector electromagnetic potential $"A"$ and complex scalar field $"\psi"$-the superconductor order parameter. The idea of Ginzburg and Landau is to formulate a theory to describe the properties of superconductors at temperatures below but close to critical one. In the case when the system possesses BEC regime with Cooper pairs formed below some temperature $T^*$ but above the criticaal one we can extend the use of the Ginzburg-Landau theory to include the Bosinic regime  Fig.\ref{(fig1)E-BEC-Sc}.

The theory is gauge invariant and it is convenient to represent the order parameter in the form $\psi(x)=\rho(x)\exp {[ie^*\theta(x)]}$,
where $\rho(x)=|\psi(x)|$ is the gauge invariant local density of Cooper pairs and to introduce the gauge invariant vector  $ Q_{k}\, = \,\partial_{k}\theta+ A_{k}$ and scalar $Q = \partial_{t}\theta+ \varphi$ fields, where $\varphi=\upsilon A_0$ is the electric scalar potential and $\upsilon$ is the speed of light in the material.  The electric $\bf E$ and magnetic $\bf H$ fields are introduced by means of electromagnetic potential $F_{\lambda\nu}\, = \, \left (\partial_{\lambda}A_{\nu}-\partial_{\nu}A_{\lambda}\right)$ in the standard way $\textbf{E}/\upsilon=(F_{01},F_{02},F_{03}$ and $\textbf{B}=(F_{32},F_{13},F_{21})$. 
 The system of equations which describes the electrodynamics of s-wave superconductors in terms of gauge invariant fields $\textbf{E},\textbf{B},\textbf{Q}$ and $Q$  is \cite{Supp,Karchev17}:
\bea
& & \overrightarrow{\nabla}\times\textbf{B}\,=\,\mu\varepsilon\frac {\partial \textbf{E}}{\partial t}-\frac {e^{*2}}{m^*}\rho^2\textbf{Q} \label{MSc101}\\
& & \overrightarrow{\nabla}\times\textbf{Q}\,=\,\textbf{B}\label{MSc102}\\
& & \overrightarrow{\nabla}\cdot\textbf{E}\,=\,\frac {\mu\varepsilon e^{*}}{D}\rho^2 \label{EScCp13}\\
& & \overrightarrow{\nabla} Q+\frac {\partial \textbf{Q}}{\partial t}\,=\,-\textbf{E}\label{MSc104}\\
& & \frac {1}{2m^*}\Delta\rho +\alpha \rho-g\rho^3- \frac {e^{*}}{D} \rho Q-\frac {e^{*2}}{2m^*}\rho \textbf{Q}^2=0,\label{EScCp15} 
\eea
where  $\mu$ is the magnetic permeability, $\varepsilon$ is the electric permittivity of the superconductor, the constant $D$ is the normal-state diffusion, ($e^*,m^*$) are effective charge and mass of superconducting quasi-particles. The constant $\alpha$ is a function of the temperature $T$
\begin{equation}\label{HSE2}
\alpha=\alpha_0(1-\frac{T}{T_{sc}})\end{equation}
where $T_{sc}$ is the superconducting critical temperature and $\alpha_0$ is a positive constant. In the temperature interval 
we consider $T_{sc}<T<T^*$ it is negative. 

The last two terms of equation (\ref{EScCp15}) are responsible for the different impact on superconductivity of applied electric and magnetic fields. If we apply magnetic field ($\textbf{E}=0,Q=0$) the qualitative analysis of equation (\ref{EScCp15}) shows that the magnetic vector potential effectively decreases the $\alpha$ parameter, $\alpha\rightarrow\alpha-e^{*2}<\textbf{Q}^2>$, where $<\textbf{Q}^2>$ is some average value. 
We can represent $\alpha^B=\alpha-e^{*2}<\textbf{Q}^2>$ in the form Eq.(\ref{HSE2}) $\alpha^B=\alpha_0(1-\frac{T}{T^B_{sc}})$, where $T^B_{sc}$ is the superconductor critical temperature in the presence of magnetic field ${\bf B}$. It is evident that the critical temperature when magnetic field is applied is lower $T^B_{sc}<T_{sc}$. 

To gain insight into the impact of the electric field on the superconductivity ($\textbf {B}=0,\textbf{Q}=0$.) we replace the scalar field $Q$ by its average value $<Q>$. The electric scalar potential effectively changes the $\alpha$ parameter
$\alpha\rightarrow\alpha-\frac {e^{*}}{D}<Q>=\alpha_r$. If $<Q>$ is positive, the applied electric field decreases the $\alpha$ parameter and destroys superconductivity. When $<Q>$ is negative the parameter $\alpha$ increases. If the system is in a normal state $T>T_{sc}$ and parameter $\alpha$ is negative, one can apply an electric field, strong enough, to change the sign of the renormalized parameter $\alpha_r>0$ which leads to Bose-condensation and onset of superconductivity. Therefore applying electric field the superconductor critical temperature increases $T^E_{sc}>T_{sc}$.

This result is fundamentally different from the result in the London brothers theory. London brothers equations imply that an electric field penetrates a distance $\lambda_L$ as a magnetic field does
\cite{London35}.

To verify the above qualitative analysis we solve numerically the system of equations for time-independent fields without magnetic
 field $\textbf {B}=\textbf{Q}=0$ and superconductors with slab geometry.
In this case the fields depend on one of the coordinates "z"  and the electric field vector has one nonzero component $\textbf{E}=(0,0,E)$. The system is in normal state $T_{sc}<T<T^*$, so that $\alpha$ is negative. 

The solutions depend on two parameters \cite{Supp,Karchev17}
\begin{eqnarray}
& & \beta\,=\, \frac {e^{*2}\mu\varepsilon}{2m^* D^2|\alpha|g}\label{HSE10} \\
& & \gamma\,=\,\frac {e^*}{D\sqrt{2m^*|\alpha|^3}} E_0 \label{HSE11}, 
\end{eqnarray}
where  $\textbf{E}_0=(0,0,E_0)$ is the applied electric field. 

\begin{figure}[!ht]
	\epsfxsize=\linewidth
	\epsfbox{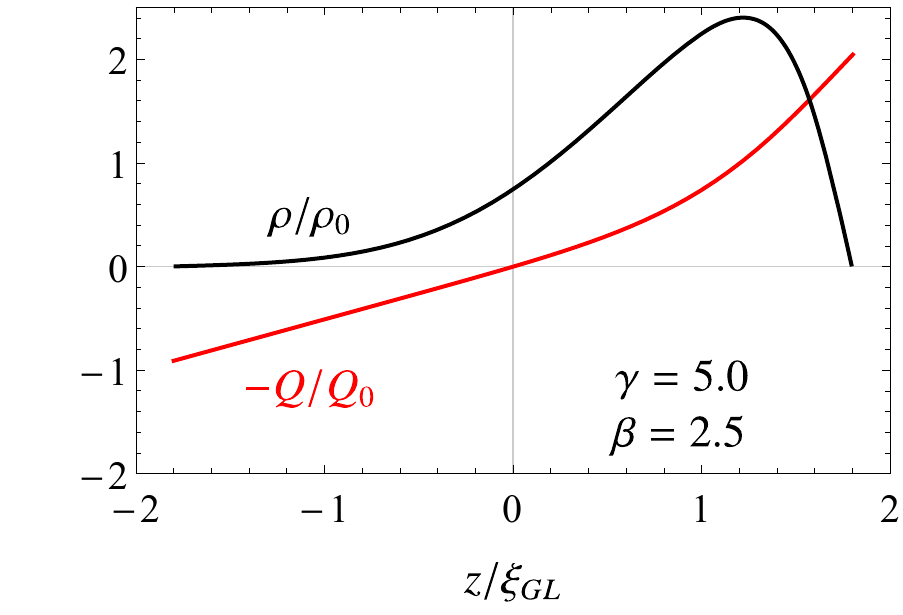} 
	\caption{\,\,The dimensionless functions $\rho/\rho_0$ and $-Q/Q_0$ of a dimensionless distance $\zeta=z/\xi_{GL}$, where $\rho_0=\sqrt{|\alpha|/g}$, $\xi_{GL}=1/\sqrt{2m^*|\alpha|}$, $Q_0= 2E_0\xi_{GL}$ and $E_0$ is applied electric field, are plotted for fixed values of	parameters 
$\gamma= 5$ and $\beta= 2.5$. For slab geometry $\zeta$ runs the interval $-1.8\leq \zeta \leq 1.8$.  The figures show an evident connection between Bose condensation of Cooper pairs $\rho$ and gauge invariant scalar field $Q$. Close to the one of the plates of the capacitor (slab geometry) the scalar potential is positive and Cooper pairs do not condense $\rho=0$. When the distance $z$ approaches the other plate $Q$ decreases and at characteristic distance $z_0$ is zero. Above it $Q$ is negative the Cooper pairs Bose condence $\rho>0$ and the system is superconductor.  }\label{(fig2)E-BEC-Sc}
\end{figure}

The dimensionless function $\rho/\rho_0$ of a dimensionless distance $\zeta=z/\xi_{GL}$  where $\rho_0=\sqrt{|\alpha|/g}$ and $\xi_{GL}=1/\sqrt{2m^*|\alpha|}$, 
and the dimensionless function $-Q/Q_0$ of $\zeta$, where $Q_0= 2E_0\xi_{GL}$, are depicted in (Fig.\ref{(fig2)E-BEC-Sc})
for fixed values of parameters $\gamma=5$ and $\beta=2.5$. We consider superconductor with slab geometry where $\zeta$ runs the interval $-1.8\leq \zeta \leq 1.8$. The figures show the relationship between Bose condensation $\rho$ and the gauge invariant scalar field $Q$. At characteristic distance $z_0$, the field $Q$ is zero. In our case, $\gamma=5$ and $\beta=2.5$, we find  $z_0=0$. When $z$ is well below $z_0$, $Q$ is
 positive and the Bose condensation of Cooper pairs $\rho$ is very small even zero. When $z$ is above $z_0$, $Q$ is negative the Cooper pairs Bose condense $\rho>0$ and the material is superconductor. 

It is important to make difference between electrons and electron pairs in electric field. The electrons expel the electric field, while the fermion pairs Bose condense. This permits us to determine $T^*$. Applying electric field above this temperature the electrons expel the electric field. The electric field, applied below $T^*$ forces the electron pairs to Bose condense, which in turn leads to increasing the superconductor critical temperature.

The calculations \cite{Supp} show that when $\beta$ is fixed $\gamma$ decreasing decreases the Cooper pair condensation $\rho$. Below critical value $\gamma_{cr}$ there is no Bose condensation of Cooper pairs. The dimensionless function $\rho/\rho_0$ as a function of dimensionless distance  $\zeta=z/\xi_{GL}$ is depicted in figure \cite{Supp} for different values of $\gamma$ and $\beta$ parameter is fixed $\beta=2.5$.
\begin{figure}[!ht]
\epsfxsize=\linewidth
\epsfbox{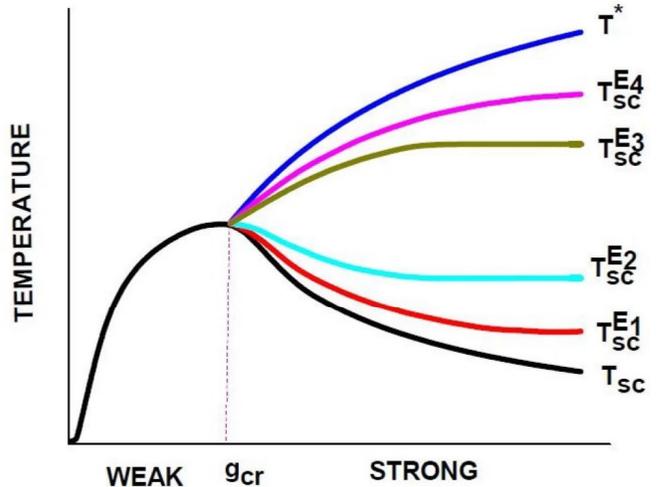} 
\caption{\,\, Schematic phase diagram of BCS-BEC crossover in the presence of applied electric field. The curve $T_{sc}$ is superconductor critical temperature as a function of attraction constant-$g$ when the electric field is not applied. The curves $T^{E_1}_{sc}$, $T^{E_2}_{sc}$, $T^{E_3}_{sc}$, and $T^{E_4}_{sc}$ are superconductor critical temperatures when electric fields $E_1<E_2<E_3<E_4$ are applied. Increasing of the critical temperature can be calculated from equation (\ref{HSE12}) if the parameters of the material are known. The curve $T^*$ shows the critical temperature of fermion pairing as a function of $g$.  $g_{cr}$ is the critical value of the BCS-BEC transition. The figure shows that if an appropriate electric field is applied, the system could possess superconductor transition at $T^*$ even though the system is in BEC regime.}\label{(fig3)E-BEC-Sc}
\end{figure}

An important consequence is the fact that an arbitrary temperature $T'$ within the interval $T_{sc}<T'<T^*$ is a superconductor transition critical temperature if an appropriate electric field is applied. To prove that we consider $T'$ from the above interval, define $|\alpha'|=\alpha_0(T'/T_{sc}-1)$ and obtain $\beta'$ from equation (\ref{HSE10}) 
replacing $\alpha$ with  $\alpha'$. The electric field $E_0'$, we have to applied, is obtained from equation (\ref{HSE11}), replacing $\gamma$ with $\gamma_{cr}$ calculated from the equations for the Bose condensation field $\rho$ \cite{Supp}: 
\begin{equation} \label{HSE12}
E_0' =  \gamma_{cr}\frac {D\sqrt{2m^*|\alpha'|^3}}{e^*}.
\end{equation}

Equation (\ref{HSE12}) provides justification for performing an experiment applying an electric field.
It serves also to prove the existence of an applied electric field that moves the system to a system with superconductor critical temperature $T^*$.
This means that
even the system is in  BEC regime, away from the BCS one, we can apply an electric field to move the system to a state with $T^E_{sc}=T^*$, characteristic of BCS regime. Above this critical applied electric field the system is metal and the electric field is expelled.

The impact of the applied electric field is illustrated graphically in figure (\ref{(fig3)E-BEC-Sc})  where schematic phase diagram of BCS-BEC crossover, in presence of electric field, is depicted.
The curves $T^{E_1}_{sc}$, $T^{E_2}_{sc}$, $T^{E_3}_{sc}$, and $T^{E_4}_{sc}$ are superconductor critical temperatures when electric fields $E_1<E_2<E_3<E_4$ are applied. Increasing of the critical temperature can be calculated from equation (\ref{HSE12}) if the parameters of the material are known. The figure shows that if an appropriate electric field is applied, the system could possess superconductor transition at $T^*$ even though the system is in BEC regime.

In conclusion, we want to emphasize that the applied electric field affects superconductivity in a way quite different from that predicted by the London brothers \cite{London35}.

Important difference is that applied electric field forces the Cooper pairs to Bose condense and in turn raises the critical temperature of the superconductor.

Inspired by this distinction we have focused on the strongly correlated superconductors which are thought to exhibit BCS-BEC crossover phenomena. Because of the interest it is important create theory and experiment for precise establishing of BCS-BEC crossover. 
Experiments with an applied electric field allow to identify the BCS-BEC crossover, in particular to distinguish $T^*$ and $T_{sc}$. Our investigation designs an experimental way for verification of the pairing of Fermions prior the superconductivity, and to measure the new characteristic temperature  $T^*$ of the system, below which the formation of Cooper pairs begins. 

We proved the existence of electric field induced superconductor transition if the system is in bosonic regime (see Fig.\ref{(fig1)E-BEC-Sc}).

Special comments on copper oxide superconductors are need. There are arguments in favor of the fermion pairs prior superconductivity, which permit to describe superconductivity in cuprates systems in evolution from Bose-Einstein regime to BCS  regime \cite{Uemura91}, and critical remarks against their formation.    
If the existence of pairs is proven applying electric field, one can decrease the critical doping of onset of superconductivity  and to raise the superconductor dome.

We emphasize that BCS-BEC crossover is not a critical point but it is characterized by interval. Within this interval one observes a significant departure from convenient BCS theory.

The results reported in this paper were obtained by examining the system of equations which
describes the electrodynamics of s-wave superconductors obtained in \cite{Karchev17}. The main mathematical result is the difference between the effects of applied magnetic and electric fields on superconductivity. As we know, magnetic field suppresses superconductivity. What is new is that the electric field leads to Bose-condensation of the Cooper pairs and to an increase in the critical temperature. We apply this technique to study two systems: in the case of H-based superconductors \cite{Karchev22} the system has a higher superconducting critical temperature if we apply an electric field instead
of increasing the pressure.
In the present paper we consider superconductors in Bose-Einsten condensation regime. We use the same mathematical technique to prove that the applied  electric field pushes Cooper pair to Bose condence and respectively to increase of the superconducting critical temperature.

\vskip 1cm

\end{document}